# Strain-magneto-optics of a magnetostrictive ferrimagnet $CoFe_2O_4$

Yu.P. Sukhorukov, A.V. Telegin, N.G. Bebenin, A.P. Nosov, V.D. Bessonov, A.A. Buchkevich

*M.N. Miheev Institute of Metal Physics of Ural Branch of Russian Academy of Sciences, Kovalevskaya str. 18, 620990 Yekaterinburg, Russia*

We experimentally demonstrate that in magnetostrictive ferrimagnetic single crystal of $CoFe_2O_4$ there is clear correlation between magnetostriction and magnetoreflection of unpolarized light in the infrared range. The influence of magnetic field on specular reflection is likely to be indirect: application of a magnetic field results in strong strain and deformation of the crystal lattice, which leads to the change in electron energy structure and hence reflection spectrum.

## I. INTRODUCTION

Optical reflection and absorption in magnetics are sensitive to magnetic field and deformation, which makes the optical measurements an effective tool for investigating electronic, magnetic, and lattice subsystems as well as interactions between them. This sensitivity is exploited in many devices.

A special part of solids is magnetostrictive materials in which application of a magnetic field gives rise to large strain and deformation because of strong coupling of magnetic moments with crystal lattice. The electronic and optical properties of these materials can be more sensitive to magnetic field comparing to solids with weak magnetostriction because magnetic field and strain work together.

Magnetooptical phenomena associated with magnetoelastic interactions are investigated by many authors (see e.g. [1, 2] and references therein). This part of the solid state physics can be called "strain-magneto-optics". The effects mentioned were studied in polarized light. It is reasonable to find whether such effects can be observed in natural (unpolarized) light because the use of natural light is much easier than polarized light. In this work, we show that magnetoelastic contribution clearly manifests itself in magnetoreflection of natural light from ferrimagnetic spinel of $CoFe_2O_4$ in the infrared (IR) spectral region. To our best knowledge, the influence of magnetostriction on magnetoreflection has not been observed yet.

We have chosen $CoFe_2O_4$ (Curie temperature is $T_C = 812$ K) as an object of investigation for the following reasons. Firstly, this spinel is an insulator and therefore the interaction of light with free charge carriers is absent. Secondly, $CoFe_2O_4$ is highly transparent in the IR spectral range. Finally, in $CoFe_2O_4$ the magnetostriction constants $\lambda_{100}$ is about $6\times10^{-4}$ at room temperature i.e. the magnetostriction of $CoFe_2O_4$ is very strong [3].

## II. SAMPLES AND EXPERIMENTAL TECHIQUE

The $CoFe_2O_4$ single crystal was grown by floating zone melting with radiation heating. The value of cubic crystal lattice parameter ($a_0 = 8.38$ Å) defined from X-ray diffraction data was found to be close to the data reported in [4]. Correspondence of chemical composition to $CoFe_2O_4$ per formula unit was confirmed by X-ray microanalysis. The electrical resistivity was found to be of above $10^5$ $\Omega\cdot$cm at room temperature. Magnetization data were obtained with the help of Lake Shore 7400 vibrating sample magnetometer. The magnetostrictive properties were characterized by strain gauge technique using the (001) oriented plate-shaped samples with in-plane typical dimensions of $10\times10$ mm$^2$ and thickness of $d = 400$ μm. The reflection coefficient $R$ measurements were carried out using the plate-shaped samples with the same crystallographic orientation but smaller typical dimensions of $4\times4\times0.22$ mm$^3$. The specular reflection coefficient was calculated as $R = I_s/I_{Al}$, where $I_s$ and $I_{Al}$ are the intensities of the unpolarized light reflected from a sample and the Al mirror, respectively. Magnetoreflection is defined here as $\Delta R/R = (R(H) - R(0))/R(0)$, where $R(H)$ and $R(0)$ are the values of reflection coefficient in an external magnetic field $H$ and in zero magnetic field, respectively. The values of $R$ and $\Delta R/R$ were measured at angles close to normal incidence of the light in the infrared spectral range from 0.8 to 30 μm at a temperature of $T = 295$ K with the relative error of 0.2%. In all cases, a magnetic field was applied in-plane to the sample surface.

## III. EXPERIMENTAL RESULTS

### A. Magnetization $M(H)$ and magnetostriction $(\Delta l/l)_{100}$.

The magnetic field dependences of magnetization and magnetostriction are shown in Fig. 1(a). At $T = 295$ K, the value of coercive field ($H_c = 80$ Oe) determined from hysteresis loops is close to one reported in [4] for high-quality single crystals of the same composition. When **H**||[100] and **H**||[010] a technical saturation is observed at $H \approx 0.9$ kOe. If magnetic field is higher the magnetization increases linearly with $H$ and reaches $M = 82$ emu/g at $H = 17$ kOe, see Fig. 1(a). This value of magnetization is close to the data reported in [4, 5].

The magnetic field dependences of magnetostriction $(\Delta l/l)_{100}$ (variation of length along the [100] axis upon magnetic field) for **H**||[100] and **H**||[010] (see Fig. 1(b)) are similar to those

reported in [3] for the unannealed $CoFe_2O_4$ single crystals. The values of $(\Delta l/l)_{100}$ at saturation for our samples exceed the values reported earlier for non-stoichiometric and doped $CoFe_2O_4$ single crystals [5-7]. For the **H**||[100] orientation, $(\Delta l/l)_{100}$ is negative. The sharp increase in the $(\Delta l/l)_{100}$ values starts at $H = 1.5$ kOe and reaches saturation (~ $-624 \times 10^{-6}$) in the applied field of $H = 3$ kOe. For the **H**||[010], the sign of $(\Delta l/l)_{100}$ is positive and parabolic growth starts immediately from $H = 0$. The saturation value of ~ $+221 \times 10^{-6}$ is reached in the field of $H = 3$ kOe. This value of $(\Delta l/l)_{100}$ is three times less than that for the **H**||[100] orientation.

For a ferromagnetic material with cubic crystal structure the relative elongation along an axis defined by the directional cosines $\beta_{x,y,z}$ can be expressed as [8]

$$\frac{\Delta l}{l} = \frac{3}{2}\lambda_{100}\left(\alpha_x^2\beta_x^2 + \alpha_y^2\beta_y^2 + \alpha_z^2\beta_z^2 - \frac{1}{3}\right) + 3\lambda_{111}\left(\alpha_x\alpha_y\beta_x\beta_y + \alpha_y\alpha_z\beta_y\beta_z + \alpha_x\alpha_z\beta_x\beta_z\right), \quad (1)$$

where $\alpha_{x,y,z}$ are the directional cosines of magnetization vector. Magnetic field is assumed to be sufficient for saturation. In our case $\alpha_z = \beta_y = \beta_z = 0$ and $\beta_x = 1$. Therefore $\lambda_{100} = -624 \times 10^{-6}$ and $(\Delta l/l)_{100}$ must be equal to $-\lambda_{100}/2$ for **H**||[010]. The experimental data presented in Fig. 1(b) indicate that sign of $(\Delta l/l)_{100}$ for **H**||[010] is indeed positive but the $(\Delta l/l)_{100}$ value is less than $|\lambda_{100}|$ not twice, but three times. Therefore, our crystal may be considered as cubic although slight distortions of cubic lattice exist.

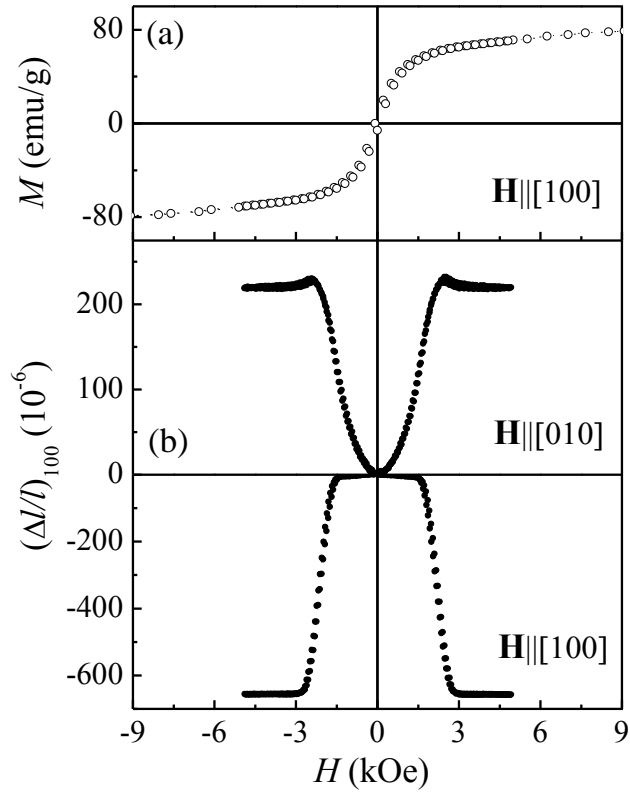

FIG. 1. Magnetic field dependences of (a) magnetization and (b) magnetostriction $(\Delta l/l)_{100}$ for the $CoFe_2O_4$ single crystal at $T = 295$ K.

## B. The spectrum of specular reflection

The spectrum of the reflection in the infrared spectral region for the $CoFe_2O_4$ single crystal at $T = 295$ K shown in Fig. 2(a) is similar to one reported for a polycrystalline sample in [9]. The spectrum of the optical conductivity $\sigma_{opt}$ calculated using the Kramers-Kronig relations from the experimental $R(\lambda)$ spectrum is plotted in Fig. 2(b). Comparison of Figs. 2(a) and (b) shows that at $\lambda < 1.5$ μm the reflection spectrum is formed by the absorption edge, by the phonon bands at $\lambda_1 = 16.4$ μm ($E_1 = 0.076$ eV) and $\lambda_2 = 24.2$ μm ($E_2 = 0.051$ eV), and by the frequency independent part of the reflection ($R \sim 14.7\%$) in the wavelength range of $1.5 < \lambda < 7.5$ μm. Weak peculiarities are seen at about $\lambda = 2 - 3$ μm. The $E_1$ band is associated with the Co-O vibrations of ions in the octahedral sublattice, while the $E_2$ band - with the vibrations of oxygen ions in the tetrahedral sublattice [10]. The long-wave edge of the $E_1$ band is distorted by additional contribution from two weak phonon bands at $\lambda_3 \approx 18.7$ μm (0.066 eV) and $\lambda_4 \approx 21.5$ μm (0.058 eV) [11]. It should be noticed that these bands are clearly seen in the calculated spectrum of optical conductivity.

## C. The spectrum of magnetoreflection

External magnetic field leads to substantial changes in reflectivity of the crystal - magnetoreflection effect. The magnetoreflection of our single crystal in the saturated field of 3.6 kOe varies from -1% to +4% depending on the spectral region. Figure 2(c) shows $\Delta R/R$ for **H**‖[100]. If **H**‖[110], the magnetoreflection is hardly detectable and is in fact within experimental accuracy. When $\lambda < 1.5$ μm, the $\Delta R/R$ grows up with decreasing $\lambda$, which is likely to be due to the shift of the absorption edge in the magnetic field. The absorption edge is known to be formed by indirect interband transitions [12]; the energy gap is somewhat higher 1 eV, which corresponds to $\lambda$ of order 1 μm. In zero magnetic field, the absorption edge shifts to shorter wavelengths with lowering temperature (so-called "blue" shift) [12, 13]. Positive sign of $\Delta R/R$ indicates that a magnetic field gives rise to the red shift of the absorption edge.

When $1.5 < \lambda < 7$ μm, there are intense peak at $\lambda = 2.96$ μm and weak maximum at about 6 μm. In the optical conductivity spectrum (Fig. 2b), the peculiarity at $\lambda = 2.96$ μm (0.42 eV) is also observed as well as in the absorption spectra of polycrystalline $CoFe_2O_4$ [14]. Similar peaks are found in spectrum of many other complex oxides. They are usually referred to as MIR (mid infrared) bands. Thus it is reasonable to suppose that the existence of the $\Delta R/R$ bands in the

$1.5 < \lambda < 7$ μm spectral range is related to variations of the intensity and position of the MIR band at $\lambda = 2.96$ μm and the fundamental edge under application of magnetic field.

When $\lambda > 7$ μm and **H**||[100] the spectral dependence of $\Delta R/R$ is characterized by the features associated with the magnetic-field-induced shift of reflection minima near the phonon bands. Near the first minimum of the reflectivity, the peculiarity is observed in narrow $\lambda$ range from 11 to 12 μm. Similar peculiarity manifests itself more clearly between the first (16.4 μm) and second (24.2 μm) phonon bands. We think that the shifts and intensity variations of two weak phonon bands at $\lambda_3(E_u$ symmetry$) = 18.7$ μm and $\lambda_4(T_{1u}$ symmetry$) = 21.5$ μm under application of a magnetic field play the most important role.

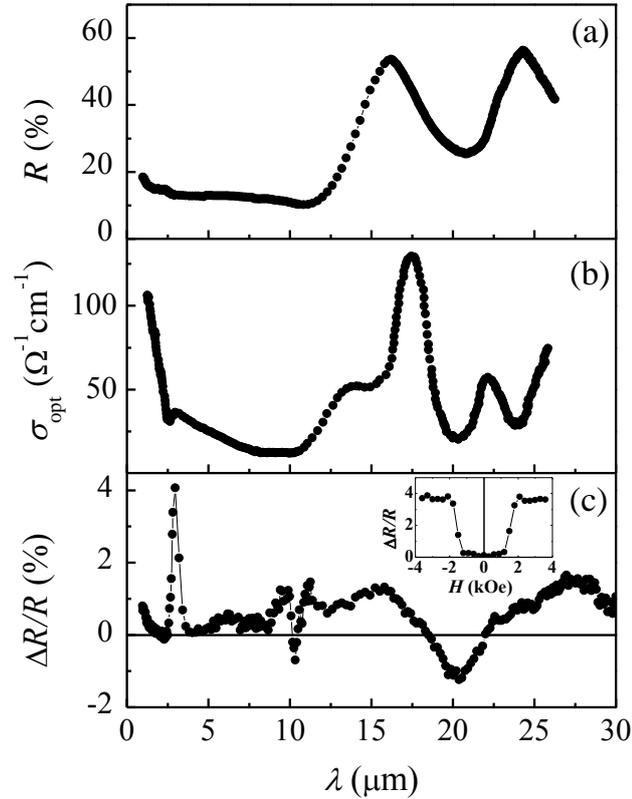

FIG. 2. The reflection spectrum $R$ at $T = 295$ K for the $CoFe_2O_4$ single crystal (a), the spectral dependence of optical conductivity $\sigma_{opt}$ calculated from the reflection spectrum using the Kramers-Kronig relations (b), and the magnetoreflection spectrum $\Delta R/R$ for **H**||[100], $H = 3.6$ kOe (c). Inset: $\Delta R/R$ vs $H$ for $\lambda = 2.96$ μm.

### III. DISCUSSION

One can see that there is close connection between magnetoreflection and magnetostriction: the magnetoreflection is large only if $(\Delta l/l)_{100}$ is large. The shape of magnetic field dependence $\Delta R/R$ (Inset in Fig.2(c)) is very similar to that of $(\Delta l/l)_{100}$. Taking into account that the red shift of the absorption edge, which is typical of magnetic semiconductor spinels like $CdCr_2Se_4$ and $HgCr_2Se_4$ [15], is not observed in $CoFe_2O_4$ we may infer that interaction of charge carriers with localized magnetic moments in $CoFe_2O_4$ is not so strong as in the spinels mentioned. Therefore, the anisotropy of optical properties of $CoFe_2O_4$ is not related to the exchange interaction of charge carrier with localized magnetic moments which occurs for $CdCr_2Se_4$ and $HgCr_2Se_4$ [15, 16]. It is more likely that the influence of magnetic field on optical properties of $CoFe_2O_4$ is indirect: magnetic field gives rise to strong strain and hence deformation of the crystal lattice which in turn results in the change of electronic spectrum. Our experiments also shows that [100] and equivalent directions play special role. This conclusion is in accordance with the results of band calculations for the $CoFe_2O_4$ [12], which indicates that the conduction band minimum lies at Γ point while the valence band tops are located at X point of Brillouin zone.

According to [12], the conduction band at Γ point and valence band at X point are non-degenerate. If the bottom of the band is situated at Γ point the shift $\Delta\varepsilon_c$ of the bottom due to deformation is known to be $\Delta\varepsilon_c = \Xi u$, where $u = u_{xx} + u_{yy} + u_{zz}$ is change in volume, $u_{ij}$ stands for deformation tensor and $\Xi$ is deformation potential. If the valence band valleys are located on [100] and equivalent axes, then $\Delta\varepsilon_v = \Xi_d u + \Xi_u u_{xx}$. When deformation is due to magnetostriction described by (1) the change in volume is absent i.e. $u = 0$, so that in sufficiently strong magnetic field, the shift of the fundamental edge is equal to $\Delta\varepsilon_v = \Xi_u \lambda_{100}$ if $\mathbf{H}\|[100]$. The magnetoreflection $\Delta R(E)/R$ can be expressed as

$$\frac{\Delta R}{R} = \frac{R(E - \Delta\varepsilon_v) - R(E)}{R(E)} \approx -\frac{d\ln[R(E)]}{dE}\Delta\varepsilon_v , \qquad (2).$$

In the vicinity of $E = 1$ eV ($\lambda \approx 1$ μm) the derivative $d\ln[R(E)]/dE$ is about 0.5 1/eV, $\Xi_u$ is usually 10 – 20 eV [17], $\lambda_{100} \approx 6.6\times10^{-4}$, therefore we obtain that $\Delta R/R$ is about 0.3 – 0.7%. The experimental value of $\Delta R/R$ at $\lambda = 1$ μm (Fig. 2) is 0.76%, so the calculated value of $\Delta R/R$ reasonably agrees with the experimental one if we take $\Xi_u = 20$ eV.

The narrow peak of magnetoreflection is found at $\lambda = 2.96$ μm. This MIR band is obviously a manifestation of deep-level impurities. Unfortunately, our data are insufficient to make a conclusion on nature of these impurities. However, the sensitivity of the peak to the magnetic field direction – and hence anisotropic strain - strongly suggest that the impurities are in a low-symmetry position.

It is stated above that when $\lambda > 7$ μm and $\mathbf{H}\|[100]$ the magnetoreflection is due to the magnetic-field-induced shift of reflection minima near the phonon bands. At first glance, such interpretation seems to be incorrect because the effect of magnetic field on crystal lattice must be extremely weak. However, in [18] it was reported that the frequencies of optical phonons in $CoFe_2O_4$ substantially depend on pressure. As the strain is controlled by the magnetic field direction because of magnetostriction, we may infer that the shift of a reflection minimum does not cause by magnetic field as such, but results from magnetic-field-induced strains.

## IV. CONCLUTION

To summarize, our study of optical properties of the $CoFe_2O_4$ magnetostrictive spinel showed that in infrared spectral range the significant magnetoreflection is observed in natural (unpolarized) light. The largest magnetoreflection has been found near MIR band where it is as high as 4%. The clear correlation has been established between magnetoreflection $\Delta R/R$ and magnetostriction $(\Delta l/l)_{100}$: large magnetoreflection is observed only if magnetostriction is great. The effect of magnetic field on optical properties of $CoFe_2O_4$ is likely to be indirect: application of a magnetic field results in strong strain and deformation of lattice, which leads to the change of spectrum – strain-magneto-optics. The deformation potential for the top of the valence band is roughly estimated as $\Xi_u = 20$ eV.

## ACKNOWLEDGMENTS

The work was conducted within the state assignment of the Federal Agency for Scientific Organizations of the Russian Federation (theme "Spin" No. 0120146330) and the Ministry of Education and Science RF (grant No. 14.Z50.31.0025).